\documentstyle[12pt]{article}

\advance\voffset -1.5cm

\def\be{\begin{equation}}\def\ee{\end{equation}}

\def\bea{\begin{eqnarray}}\def\eea{\end{eqnarray}}

\begin{document}

\noindent

\vskip 1.5cm

\begin{center}

{\large\bf
Mean Field Theory for Underdamped Josephson Junction Arrays with an
Offset Voltage
              }\\
\vskip 1.0cm

Eric Roddick and D.Stroud

\vskip .2cm

{\it
Department of Physics,
The Ohio State University, Columbus, Ohio 43210

}\\

\vspace{1.5cm}

{\bf ABSTRACT} \\

\begin{quotation}
We study a model Hamiltonian for superconductivity in underdamped Josephson
junction arrays in the presence of an offset voltage between the array and
the substrate.  We develop an approximate zero-temperature (T = 0) phase
diagram
as a function of Josephson coupling, charging energy, and offset voltage,
using a simple Hartree-type mean-field approximation.  With diagonal
charging energy, the calculated phase diagram is periodic in offset
voltage, in agreement with previous results.  At a special value of this
voltage such that states with n and n+1 Cooper pairs per grain are
degenerate, only an infinitesimal Josephson coupling is needed to establish
long-range phase coherence in this approximation.  With both diagonal and
nearest-neighbor charging energies, the T = 0 phase diagram has two
types of insulating lobes with different kinds of charge order, and two
types of superconducting regions.  One of  these is a ``supersolid'' in which
long-range phase coherence coexists with a frozen charge density wave.
We briefly discuss connections to previous calculations, and possible
relevance to experiments.


\end{quotation}
\end{center}
\vskip 1.5cm

\section{Introduction}

Josephson junction arrays have been the subject of considerable recent
research.\cite{expov}
Such arrays consist of superconducting (S) grains embedded in a
nonsuperconducting host and coupled together by the Josephson effect or by
proximity tunneling.  They exhibit a large variety of unusual behavior, both
static and dynamic, in the presence of applied d. c. and a. c. currents and
perpendicular magnetic fields.   In general, the nonsuperconducting
component can be a normal metal (N) or an insulating layer (I).  While most
experimental work has been carried out on the former \cite{expov}, SI arrays
exhibit a number of new phenomena.  For example, the dynamical properties of
such arrays are characterized by a McCumber-Stewart parameter $\beta \gg 1$,
corresponding to coupled highly underdamped Josephson
junctions.  Under such conditions, one observes hysteresis and
resistance steps in
the IV characteristics \cite{vander1,vander2},
which are reproduced by calculations.
\cite{yu1,yu2,lobb}  Vortices
in such arrays have been reported to move ballistically - that is, they
may behave like massive objects which can maintain their motion,
once initiated, even in an external driving current.\cite{vander1,rzchow}

When the S grains are sufficiently small, the behavior of an array is
modified by quantum effects.  These effects arise
from the noncommutativity of the Cooper pair number operator and the
phase of the superconducting wave function.  Such effects
were first discussed by
Anderson,\cite{and} Abeles,\cite{abe} and Simanek\cite{sim1}, and have since
been studied by a number of
authors.\cite{sim4,doniach,jose1,jacobs}  For sufficiently small
grains, quantum
phase fluctuations lead to the suppression of superconductivity entirely.
\cite{sim4,doniach,jose1,jacobs,sim2,sim3,woo,fishma,fisher88}
The array is instead in an insulating state at T = 0, in which there is
no phase ordering and the average number of Cooper pairs on each grain is
fixed.  The interactions between the charges themselves can be
logarithmic in the insulating state\cite{fazio91}.
The phase transition has also been
found to be significantly affected by dissipative
tunneling of single
electrons.\cite{zwerger,fazio90,chakravarty86,chakravarty87,kampf}

In this paper, we discuss the zero-temperature behavior of an underdamped
array in the presence of an additional control parameter: the offset
voltage between the array and a substrate plane.  Such an offset voltage
behaves like a chemical potential for injection of Cooper pairs into the
array.  As has been discussed by several workers,\cite{fisher,bruder}
it can have a complex
effect on the phase diagram of the array.  Depending on the range of the
charging interaction, one can even see evidence of ``charge frustration''
analogous to the better-known effects of a transverse magnetic field in an
overdamped array.

Our contribution here is to describe a simple mean-field
approximation for the phase diagram of an SIS array in the
presence of an offset voltage.  The approximation is a straightforward
extension of previous approaches at zero offset voltage.
\cite{sim1,sim2,sim3,woo,fishma}  It is readily
tractable, yet leads to a complex phase diagram with a rich variety of
possible phases.  In several instances, it agrees well with previous
calculations based on exact statements about the Hamiltonian.   In part of
the phase diagram, we find a ``supersolid''\cite{bruder,matsuda,liu}
phase in which superconducting order is found to coexist with a frozen
charge-density-wave.  While such a supersolid phase has been reported
previously in a similar model\cite{bruder},
it was found only in a narrow sliver of parameter
space and only with longer-range interactions than ours.

We turn now to the body of the paper.
In Section II, we describe the model Hamiltonian, as well as the simple
approximation used to treat it.  Section III presents calculated phase
diagrams at T = 0 for several ranges of parameters.  Finally, in Section
IV, we compare our results to previous calculations, describe possible
extensions, and discuss the relevance of the results to possible
experiments on artificially synthesized arrays as well as to granular
systems.

\section{Model Hamiltonian and Mean-Field Approximation}

We consider a two-dimensional array of N superconducting grains
separated from a substrate by a thin insulating layer.  The i$^{th}$ grain
is described by a superconducting order parameter
$\psi_i \equiv |\psi_i|e^{i\phi_i}$ and contains n$_i$ Cooper
pairs of charge 2e.  The i$^{th}$ and j$^{th}$ grains are coupled
by a Josephson junction with critical current
I$_{c;ij} \equiv 2eJ_{ij}/\hbar$.

We approximate the behavior of the array by the following Hamiltonian:
\begin{eqnarray}
  H=\frac{1}{2}\sum_{ij}U_{ij}n_{i}n_j + 2eV\sum_{i}n_i
    + \nonumber \\
+ \frac{1}{2}\sum_{ij}J_{ij}
  (1-\cos(\phi_{i}-\phi_{j})).
\end{eqnarray}
where the sums run over all grains i and j.
The first summation represents the Coulomb interaction among the Cooper
pairs on the various superconducting islands.  The second term is the
potential energy generated when the superconducting array is held at a
potential V with respect to the common ground.  Finally,
the last term represents the energy of
Josephson coupling between neighbouring grains.
We assume a periodic lattice of identical grains.   This implies
that J$_{ij}$ and U$_{ij}$ are functions only of the separation
$|\vec{R}_i-\vec{R}_j|$ between grains i and j.

The first summation can also be written in the form
\begin{equation}
\frac{1}{2}\sum_{ij}n_iU_{ij}n_j = \frac{1}{2}\sum_{ij}V_iC_{ij}V_j
\end{equation}
where V$_i$ is the potential on the i$^{th}$ superconducting island, and
C$_{ij}$ is the capacitance matrix.  This makes clear that
U$_{ij}$ = 4e$^2$(C$^{-1})_{ij}$, i. e., the matrix U is proportional to
the inverse capacitance matrix.

To within a constant factor, the
Hamiltonian (1) can also be written in the form
\begin{equation}
H = \frac{1}{2}\sum_{ij}(n_i-\bar{n})U_{ij}(n_j-\bar{n})
+ \frac{1}{2}\sum_{ij}J_{ij}(1-cos(\phi_i-\phi_j)),
\end{equation}
where $\bar{n}$ is a constant which is determined by the offset
potential V.
It is readily established that forms (1) and (3) differ only by the
constant term $\frac{1}{2}\sum_{ij}\bar{n}U_{ij}\bar{n}$.
Henceforth, we use the form (3), which is more convenient for
calculation.  We also assume that the coupling energy J$_{ij}$ vanishes
except for nearest neighbor grains, for which it has value J;
and that the energy U$_{ij}$ vanishes except for diagonal and nearest
neighbor contributions, which have values U$_0$ and U$_1$ respectively.

In order to motivate the mean-field approximation, we consider the types of
order which can be expected from this model Hamiltonian.  If both U$_0$ and
U$_1$ are zero, the array will undergo a transition to a state of long-
range phase coherence below a critical temperature T$_c$.
\cite{kos1,kos2}  For a square
array, it is known that T$_c \approx 0.95J/k_B$, and the transition is in
the Kosterlitz-Thouless universality class.\cite{kt2d}
The low temperature phase is
superconducting but has a novel type of long-range phase coherence in which
the phase correlation functions decay algebraically.  When
U$_0$ and U$_1$ are finite, the phase-ordering transition temperature is
reduced, reaching zero at critical values of U$_0$ and U$_1$.
At larger
values of these parameters, one expects some kind of {\em charge} ordered
state, the exact nature of which will depend on U$_0$, U$_1$, and
$\bar{n}$.

There are various ways to develop a mean-field approximation for this
Hamiltonian.   One approach is to decouple the terms which
involve more than one grains as follows:
\begin{eqnarray}
(n_i-\bar{n})(n_j-\bar{n}) \approx \frac{1}{2}(n_i-\bar{n})(<n_j>-\bar{n})
+ \frac{1}{2}(<n_i>-\bar{n})(n_j-\bar{n}) \\
cos(\phi_i-\phi_j) \approx \frac{1}{2}(cos\phi_i<cos\phi_j>+<cos\phi_i>
cos\phi_j+ \nonumber \\
+ sin\phi_i<sin\phi_j>+<sin\phi_i>sin\phi_j).
\end{eqnarray}

In effect, this
approximation converts the many-body Hamiltonian H into a sum of
single-body Hamiltonians.  The values of the canonical averages $<n_i>$,
$<cos\phi_i>$ and $<sin\phi_i>$
are then determined self-consistently by the following
procedure.  First, one makes an initial guess for the values of these
quantities.  Next, given these initial guesses, one calculates the
eigenstates of H$_{MF}$, which is the approximation to H resulting
from the substitutions (4) and (5).   Since H$_{MF}$ is a sum of
single-particle
operators, each such eigenstate is a product of single-particle
eigenstates $\psi_{i}(\phi_i)$, which are solutions to
the appropriate single-particle Schr\"{o}dinger equation:
\begin{eqnarray}
\{U_{0}(n_i-\bar{n})^2 + U_1(n_i - \bar{n})\sum_{j}\, ^{\prime}(<n_j>-\bar{n})-
\nonumber \\
- \sum_j\,^{\prime}J(<cos\phi_j>cos\phi_i
+<sin\phi_j>sin\phi_i)\}\psi_{i}(\phi_i)
=E_{i}\psi_{i}(\phi_i),
\end{eqnarray}
where the primes indicate that sums are to be carried out over all nearest
neighbors to the site i.
Eq. (6) can be expressed as an ordinary differential equation with the
help of the representation n$_i = -i\frac{d}{d\phi_i}$.
This expression follows from the canonical conjugacy of
the charge operator n$_i$ and the phase $\phi_i$,
which implies the commutation relation
$[n_i, \phi_i]$ = -i.\cite{and,abe,sim1}

In seeking self-consistent solutions to eq. (6), we have made the
assumption that $<\sin\phi_i> = 0$, i. e., that the phase order parameters
of all the grains are parallel, although we do allow for the possibility
that the \underline{amplitudes}
of these order parameters are unequal on different
sites.   With this assumption, and writing
\begin{equation}
\psi_{i}(\phi_{i})=\exp(i\eta_i x_i)f_{i}(x_i),
\end{equation}
with $x_i = 2\phi_i$ and
$\eta_i=\bar{n}-\frac{U_{1}}{U_{0}}\sum_j^{\prime}(<n_{j}>-\bar{n})$,
we can express the Schr\"{o}dinger equation in the form of Mathieu's equation:
for f$_i$:
\begin{equation}
f_{i}^{\prime\prime}(x_i) + (e_i + 2q_i\cos2x_i)f_i(x_i)=0,
\end{equation}
where $e_i=2E_{i}/U_{0}$ and $q_i=\sum_j^{\prime}J<\cos \phi_{j}>/U_{0}$.
Note that since $\phi_i$ is the phase of the superconducting order parameter on
the i$^{th}$ site, the solutions $\psi_{i}(\phi_i)$
must be 2$\pi$ periodic. This means $f_i(x_i+\pi)=\exp(i\eta_i\pi)f(x_i)$.

Given the single-particle solutions to eq. (6), the
values of $<n_{i}>$ and $<\cos\phi_i>$ at
temperature T can be
calculated self-consistently from the relation:
\begin{eqnarray}
   <n_{i}> &=& \frac{\sum_{\nu}\exp(-\beta E_{\nu})
       <\Psi_{\nu}|n_{i}|\Psi_{\nu}>}{\sum_{\nu}\exp((-\beta E_{\nu})}
\nonumber \\
  <\cos\phi_{i}> &=&\frac{\sum_{\nu}\exp(-\beta E_{\nu})
               <\Psi_{\nu}|\cos \phi_{i}|\Psi_{\nu}>}
      {\sum_{\nu}\exp(-\beta E_{\nu})}
\end{eqnarray}
where $\beta =1/k_BT$, $\Psi_{\nu}$ denotes the product of the single-particle
wave-functions, and E$_{\nu}$ is the corresponding energy eigenvalue, which
is the sum of single-particle eigenvalues.  Because of the Hartree
approximation, these averages simplify to involve sums over
only single-particle states on the i$^{th}$ grain.
Once the averages are calculated,
they are substituted back into the
single-particle Schr\"{o}dinger
equations (6), and the process is repeated until
convergence is obtained.

In the present paper, we will be concerned primarily with the phase diagram
at T = 0.   The self-consistency procedure is then considerably simplified,
since only the ground state wave functions enter into the sums in (9).
The required equations reduce to
\begin{eqnarray}
<\cos\phi_{i}>_{o} = <\Psi_{o}|\cos\phi_{i}|\Psi_{o}> \nonumber \\
<n_{i}>_{o} = <\Psi_{o}|n_{i}|\Psi_{o}>,
\end{eqnarray}
where the subscript $o$ refers to the ground state.

\section{Results.}

We have carried out calculations for a bipartite lattice (i. e., one which
can be decomposed into two sublattices), assuming only nearest neighbor
couplings J$_{ij}$ which we denote J.  We seek self-consistent
solutions in which $<n_i>$ and $<\cos\phi_i>$ can each
have different values on the two sublattices.
The resulting phase diagram depends on the
parameters $\bar{n}$, U$_0$, U$_1$, and J.  Since
only the ratios of the last three parameters enter nontrivially, we
introduce dimensionless parameters $\alpha = zJ/U_0$ and
$\gamma = zU_1/U_0$, where z is the number of nearest neighbors.  Thermal
effects are described in terms of a scaled temperature k$_B$T/U$_0$.

We begin by considering the case of only diagonal charging energy, i. e.
$\gamma = 0$.   In the special case $\bar{n} = 0$, this problem has been
previously treated by a number of workers.
\cite{sim1,sim2,sim3,woo,fishma}
For any given values of $\bar{n}$ and $\alpha$,
the self-consistently determined value of
$<cos \phi_i>_0$ in the ground
state is found to be independent of i - that is, the ground state is
translationally invariant.  As $\alpha$ decreases,
$<cos \phi>$ also diminishes, reaching zero at a critical value
$\alpha_c(\bar{n})$.  At this point
phase fluctuations induced by the charging energy become strong enough to
destroy superconducting order.
Fig. 1 shows $<cos\phi>$ as a function of $\alpha$
for several values of $\bar{n}$.  The dependence of $\alpha_c$ on
$\bar{n}$ is shown in Fig. 2.

The behavior shown in Fig. 2 is readily understood.  The
Hamiltonian (1) is periodic in $\bar{n}$ with period unity.  This implies
that the phase diagram should be similarly periodic, repeating at each
integer number of Cooper pairs.  This periodicity has previously been noted
by Bruder {\it et al} \cite{bruder} on the basis of general consideration.
Fisher {\it et al} \cite{fisher} also obtained such a periodicity
in the context of a somewhat different model involving bosons with a
short-range repulsive interaction
on a periodic lattice.  Besides this periodicity,
the symmetry of the Hamiltonian implies that
$\alpha_c(\bar{n})=\alpha_c(1-\bar{n})$, a symmetry which is also reflected
in Fig. 2.
The critical value $\alpha_c$ has its minimum value at
$\bar{n} = 1/2$, where $\alpha_c = 0$.   Thus, at $\bar{n} = 1/2$, a
superconducting ordered state
can be established with even an infinitesimal Josephson
coupling, at least within the mean-field approximation.

To better understand the behavior near half-integer $\bar{n}$, we note that
the ``interaction term,'' H$_{int}$ = -2J$<cos\phi>$cos$\phi_i$,
behaves like a perturbation on the ``kinetic energy term''
H$_0=\frac{1}{2}U_0(-i\frac{d}{d\phi}-\bar{n})^2$ in the Schr\"{o}dinger
equation for $\psi_i(\phi_i)$.  In the absence
of this perturbation,
the states with n and n+1 pairs are
degenerate at $\bar{n}=n+\frac{1}{2}$ (n = integer),
both representing the ground state of the unperturbed
Hamiltonian (cf. Fig. 3a, where this degeneracy is depicted for the
special case $\bar{n} = \frac{1}{2}$).
The interaction term breaks this degeneracy and produces phase ordering.
We may estimate $<cos\phi>$ in the mean-field
approximation at finite temperatures, including only these two lowest
states in the canonical average.  The resulting self-consistent equation
for $<\cos\phi>$ at half-integer $\bar{n}$ is
\begin{equation}
  <\cos \phi>=\frac{2}{4-\alpha}\tanh(\frac{\alpha U_0}{k_{B}T}<cos\phi>).
\end{equation}

Fig. 3(b) shows the temperature-dependent
phase order parameter $<\cos\phi>$ resulting from (11).
For k$_BT \ll U_0$, $<\cos \phi>$ goes to zero at
$\alpha=2k_{B}T/U_0$, or $T = zJ/(2k_B)$.  Thus, for any
finite J, $<cos\phi>$ remains finite up to this temperature, irrespective
of the value of the charging energy U$_0$.  The physics
behind this behavior is easily understood.  Precisely at half-integer
$\bar{n}$, the two degenerate states are split by the phase-coupling
perturbation into two states, in each of which the phase order parameter
is nonzero.
The energy splitting is of order zJ/2.  As long
as k$_BT$ is smaller than this value, the lower state is predominantly
occupied, leading to a nonzero phase order parameter.  When k$_BT$ exceeds
this value, $<\cos\phi>$ drops to zero.

We turn next to the mean-field phase diagram when the off-diagonal charging
term $\gamma \neq 0$.
Figs. 4 and 5
show the zero-temperature mean-field phase diagram
for two values of $\gamma$ in the range
$0<\gamma<1$.\cite{gamma}
The phase diagram is considerably more complex than for
$\gamma = 0$.  At small values of $\alpha$, the system is in a
nonsuperconducting phase, with $<cos\phi_i>=0$, but we can identify
two different types of charge structure.  Near integer $\bar{n}$, we
have a ``ferromagnetic'' charge ordering, such that $<n_i>$ equals the same
integer on each grain.  Near half-integer $\bar{n}$, the charge ordering
is ``antiferromagnetic'', provided the lattice is bipartite.
In this case, $<n_i>$ takes on two
distinct integer values on the two sublattices,
say $<n_i> = n$ and $<n_i> = n+1$, with n integer.  These phases are
``incompressible'' in the sense that the total number of Cooper pairs on
the lattice does not vary continuously with $\bar{n}$, but rather jumps
discontinuously from one value to another at the phase boundary.  These
phases are thus ``Mott insulators'' as previously noted by Fazio and Sch\"{o}n.
As the value of $\gamma$ increases, Figs. 4 and 5
show that the ``antiferromagnetic'' lobe expands at the expense of the
``ferromagnetic'' lobes.  This is to be expected, as no antiferromagnetic
lobe exists when $\gamma = 0$.

The mean-field approximation predicts that
superconducting phase, which occurs at large $\alpha$, also
coexist with two different types of charge structure, depending on the
value of $\bar{n}$.  Near $\bar{n} = 1/2$ we find an antiferromagnetic
charge structure coexisting with a nonzero $<cos\phi_i>$ (corresponding to
a superconducting phase).  In this phase, both $<n_i>$ and
the phase order parameter $<cos\phi_i>$ takes on two
different values on the two grain sublattices.  In contrast to the
corresponding Mott insulating phase, however, $<n_i>$ is always a
\underline{continuous} function of $\bar{n}$ in the superconducting
phase.  Because of this, we expect that the phase boundary between
the charge-ordered and non-charge-ordered superconducting phases may
also correspond to a continuous, rather than first-order, phase transition.
As in the
nonsuperconducting phases at small $\alpha$, the charge-ordered
superconducting lobe shrinks as $\gamma$ becomes smaller.

\section{Discussion}

It is of interest to compare our results with recent calculations by Fisher
{\it et al}$\cite{fisher}$ and by Bruder {\it et al} $\cite{bruder}$,
based on related
models.  Fisher {\it et al} consider
a model similar to ours, but with $\gamma = 0$.
They also solve their model hamiltonian in
a mean-field approximation, which is presented as an exact solution to
their model in the limit of large $\bar{n}$ and infinite-range hopping.
Despite these differences, their phase diagram
at $T=0$ closely resembles ours for the $\gamma=0$ case, with
insulating lobes centered at each integer $\bar{n}$.

Bruder {\it et al} $\cite{bruder}$ consider a model Hamiltonian identical
to ours  but, rather than solving it in a
mean-field approximation, they map it onto the so-called
XXZ spin-1/2 Heisenberg model in two dimensions.  This model is equivalent
to ours only in the limit of ``hard-core bosons,'' i. e. U$_0 = \infty$.
In this limit, the occupancy of a given grain is restricted to either zero
or one Cooper pair.
With only nearest-neighbor coupling, they do not
obtain one striking result of our calculations,
namely, a charge-ordered state
which coexists with superconductivity.  This appears in their approximation
only when next-nearest-neighbors are included, and even then only
in a narrow region of their phase diagram.   This sliver occurs at
small $\alpha$, and
never includes half-integer $\bar{n}$ as does our supersolid
phase.

The phase diagram of
Bruder {\it et al} is very similar to ours in
other respects, with insulating
lobes centered at half-integer and integer $\bar{n}$ (for the case of
diagonal and nearest-neighbor charging energy).
At precisely half-integer $\bar{n}$ they
find a phase transition between a charge-ordered insulating state and a
non-charge-ordered superconducting phase at the so-called Heisenberg
point:  $(J/U_{1})_{cr} = 1$.  This value
is in excellent agreement with our results in the limit of
large $U_{0}$.  For example, when $\gamma = 0.3$, we observe a phase
transition at $\alpha = 0.26$,  corresponding to
$(J/U_{1})_{cr} = 0.26/0.3 = 0.87$.  As $U_0$ increases, this ratio
more and more closely approximates the Heisenberg value.  Thus,
for $\gamma=0.2$, the transition
occurs at $\alpha = 0.19$, $(J/U_1)_{cr}=0.95$,
while for $\gamma=0.1$, it takes place at
$\alpha=0.1$, $(J/U_1)_{cr}=1.0$.  Thus,
our mean-field
approach agrees reasonably well with available
exact results in the appropriate limits.

The  ``supersolid'' phase, corresponding to the coexistence of charge and
superconducting order, has a simple interpretation in
the language of the ``XXZ'' model: the phase
order represents ferromagnetic ordering in the xy plane, while charge
ordering represents
antiferromagnetic order in the z direction.\cite{matsuda}  It
is not clear how such a state could be detected experimentally, but since
it is presumably zero-resistivity, the
charge-ordered state could presumably be set in motion with an
arbitrarily small applied voltage.  This might lead to unusual effects
involving oscillating currents in the superconducting state.

Another result that emerges from our work, as well as that of
Sch\"{o}n and coworkers, is the existence of superconductivity at
arbitrarily weak Josephson coupling at half-filling ($\bar{n}=1/2$) in the
case of diagonal capacitance energies.  It is amusing to note that in
the alkali-doped fullerenes, superconductivity is most conspicuous in
materials (e. g. K$_3$C$_{60}$) which have half-integer numbers of Cooper
pairs per ``grain'' (i. e. the C$_{60}$ clusters).  If the C$_{60}$ lattice
could be viewed as Josephson-coupled superconducting grains, described by
the Hamiltonian (1), then the maximum T$_c$ would occur at half-integer
Cooper pair number.  However, it seems most unlikely that such small
islands could appropriately be viewed as Josephson-coupled grains.

It would be straightforward to extend the present calculations to the
case of randomness in both the capacitance and the offset potential.
Fisher {\it et al}\cite{fisher} have considered the effects
of a random external potential on their interacting Bose model.
For the case of weak bounded disorder, they find that the superfluid
lobes are narrowed,
and a new phase emerges: an insulating, gapless Bose glass.
They argue that on site randomness plays a crucial role in the phase diagram of
$^4$He adsorbed in porous media.  We expect that our mean-field theory
would also yield a Bose-glass-like phase in an appropriate part of the
phase diagram.  This can be seen by considering, e. g., the case of random
diagonal capacitance energies.  In the insulating regime, for a continuous
bounded disorder, we expect that, in the mean-field approximation,
the charge number should vary continuously
with offset potential in certain ranges of this potential, as successively
more grains jump from one charge state to another.  This would correspond
to a ``compressible, insulating'' phase, similar to the Bose glass regime
of Fisher {\it et al}.  Presumably the ``Mott'' (incompressible) phase would
disappear altogether for sufficiently strong disorder.

As far as we know, there are no published experiments to date
which show the lobe structure described in this paper.  Similar
conductance oscillations are, however, well-established in
small groups of normal junctions.\cite{normal}  Presumably, it
would be straightforward to carry out such measurements with an underdamped
array in an appropriate geometry, seeking periodic variations in
conductance with offset voltage.  It would certainly be of great
interest of such oscillations could be observed in superconducting
arrays.

\section{Acknowledgments.}  We are grateful to M. Makivic for
valuable conversations.  This work has been supported by the National
Science Foundation, grant DMR90-20994.

\newpage
\begin{center}
{\bf Figure Captions}
\end{center}

\begin{enumerate}

\item
Plot of the phase order parameter $<cos\phi>$ as a function of the
parameter $\alpha \equiv zJ/U_0$ at temperature T = 0, for several
values of the average Cooper pair number $\bar{n}$, as calculated in the
mean-field approximation.  The nearest neighbor charging energy U$_1$ =0
in this calculation.

\item
Plot of the critical charging energy parameter $\alpha_c$ as a function
of $\bar{n}$ for the case of diagonal charging energy only, as calculated
in the mean-field approximation.

\item
(a). Plot of the ground state energy E$_0$ as a function of $\bar{n}$ for
the case of diagonal charging energy only, and zero Josephson coupling
energy J.  The dashed lines indicate how the doubly-degenerate ground-state
energy is split near $\bar{n} = 1/2$ in the presence of a small Josephson
energy J.  (b) Plot of the phase order parameter $<cos\phi>$ as a function
of temperature at $\bar{n} = 1/2$, for small values of $\alpha = zJ/U_0$.
The phase order parameter drops to zero in this limit near k$_BT=zJ/2$.

\item
Phase diagram in the $\alpha$-$\bar{n}$ plane at T = 0 for the case
$\gamma \equiv zU_1/U_0 = 0.5$, as calculated in the mean-field
approximation.  The different types of order in the ground state are
indicated by the legends in the Figure.

\item
Same as Fig. 4, but for $\gamma=0.4$.

\end{enumerate}

\setcounter{equation}{0}



\begin{thebibliography}{99}
\bibitem{gamma}
  Note that only certain values of the ratio $\gamma$ are
physically acceptable.  For example, if $\gamma > 1$, the Hamiltonian
has no lower bound, and there does not exist a ground state energy.
This also follows from the fact that both U$_0$ and U$_1$ are
appropriate matrix elements of the inverse capacitance matrix for the
array.  If these elements are explicitly calculated,
it can readily be shown that the ratio $\gamma$ cannot exceed unity.
\end{thebibliography}
\end{document}